\documentclass{iau} 
\usepackage{graphicx}

\newcommand{\msun}{{\rm M}_\odot}
\newcommand{\Msun}{M_\odot}

\newcommand{\ifm}[1]{\relax\ifmmode#1\else$\mathsurround=0pt #1$\fi}
\newcommand{\kms}{\ifmmode\,{\rm km}\,{\rm s}^{-1}\else km$\,$s$^{-1}$\fi}

\newcommand{\kpc}{\,{\rm kpc}}

\newcommand{\GyrI}{\,{\rm Gyr}^{-1}}

\def\sy{\,M_\odot\, {\rm yr}^{-1}}

\def\Ms{M_*}

\title[Simulations at the Dwarf Scale] 
{Simulations at the Dwarf Scale: From Violent Dwarfs at Cosmic Dawn and Cosmic Noon to Quiet Discs today}

\author[Daniel Ceverino]   
{Daniel Ceverino$^1$}

\affiliation{$^{1}$Universitat Heidelberg, Zentrum für Astronomie, Institut fur Theoretische Astrophysik, Albert-Ueberle-Str. 2, 69120 Heidelberg, Germany}

\pubyear{2019}
\volume{xxx}  
\jname{Title of your IAU Symposium}
\editors{A.C. Editor, B.D. Editor \& C.E. Editor, eds.}
\begin{document}

\maketitle

\begin{abstract}
Dwarf galaxies with stellar masses around $10^9 \Msun$ can be explored at high and low redshifts and they give a glimpse of the different conditions of galaxy formation at different epochs. Using a large sample of about 300 zoom-in cosmological hydrodynamical simulations of galaxy formation I will briefly describe the formation of dwarfs at this mass scale at 3 different epochs: cosmic dawn (\cite[Ceverino, Klessen, Glover 2018]{paperII}), cosmic noon (\cite[Ceverino, Primack, Dekel 2015]{Ceverino et al. 2015b}), and today (\cite[Ceverino et al. 2017]{Ceverino17}). 
I will describe the FirstLight simulations of first galaxies at redshifts 5-15. These first dwarfs have extremely high star formation efficiencies due to high gas fractions and high gas accretion rates. These simulations will make predictions that will be tested for the first time with the James Webb Space Telescope (JWST).  
At cosmic noon, $z=2$, galaxy formation is still a very violent and dynamic process. The VELA simulations have generated a set of dispersion-dominated dwarfs  that show an elongated morphology due to their prolate dark-matter halos. 
Between $z=1$ and 0, the AGORA simulation shows the formation of a low-mass disc due to slow gas accretion. The disc agrees with many local scaling relations, such as the stellar-mass-halo-mass and the baryonic Tully-Fisher relation.
\end{abstract}


In this talk, we are going to use the Universe as a time machine. We are focusing on bright dwarf galaxies at the LMC scale, a stellar mass of a few $10^9 \Msun$. At this fixed mass, we are going to see how galaxies look like at three different epochs in the history of the Universe, using cosmological zoom-in simulations of galaxy formation (\cite[Ceverino et al. 2014]{Ceverino14}).
The simulation at $z=0$ is using the AGORA initial conditions (\cite[Ceverino et al. 2017]{Ceverino17}). 
The results at cosmic noon, $z\simeq2$, are coming from the VELA simulations (\cite[Ceverino, Primack, Dekel 2015]{Ceverino et al. 2015b}).
Finally the results at cosmic dawn, $z\geq6$, are using the FirstLight simulations (\cite[Ceverino, Glover, Klessen 2017]{paperI}; \cite[Ceverino, Klessen, Glover 2018]{paperII}).

\section{Isolated Dwarfs today ($z=0$)}

\begin{figure}[b]
\begin{center}
 \includegraphics[width=3.4in]{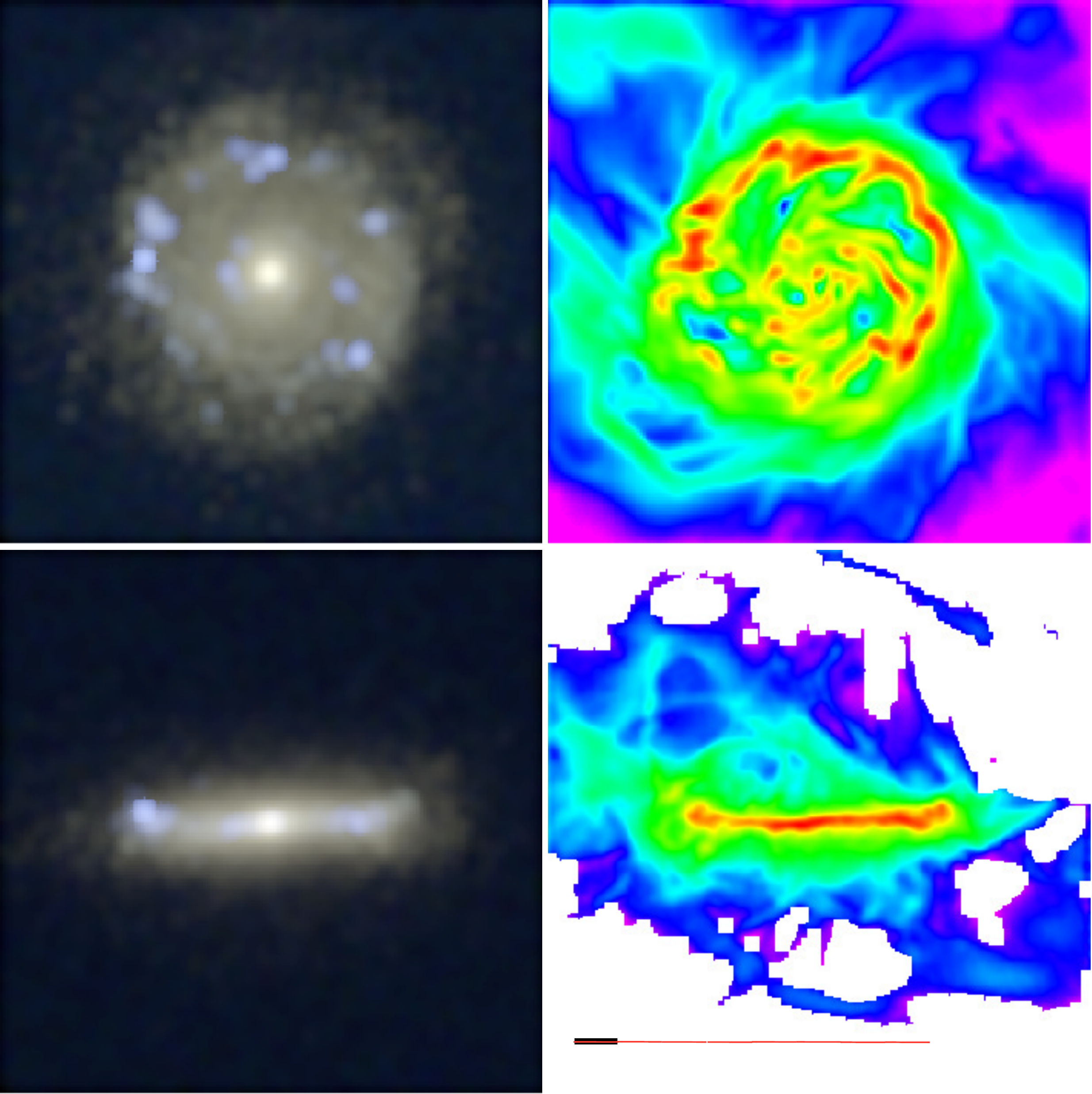} 
 \caption{UBV stellar light (left) and gas (right) of the disc galaxy at $z=0.1$ in the face-on (top) and edge-on (bottom) views (40x40 kpc), according to the direction of the angular momentum of the cold ($T<10^4 K$) gas. The horizontal bar in the bottom-right panel represents a length of 3 kpc.}
   \label{fig1}
\end{center}
\end{figure}

The galaxy shows a disc morphology (Fig.\,\ref{fig1}) both in stellar light and gas.
The disc extends to 10 kpc with a total stellar mass of $\Ms=3.2 \times 10^9 \Msun$. This is the regime of low-mass late-type galaxies.
The gas looks clumpy and flocculent, similar to the HI distribution of galaxies with similar mass.
The star-forming regions appear in the U-band and they coincide with large gas complexes, mostly organised in a star-forming ring.
The star-formation-rate, $ {\rm SFR}=0.2 \sy$, is consistent with observations of star-forming galaxies with a similar mass in the local Universe.

The stellar-to-halo mass ratio, or galaxy efficiency, is consistent with current abundance matching models, within the typical uncertainties. The simulation also agrees remarkably well with the baryonic Tully-Fisher relation.
The flat rotation velocity is estimated as the rotational velocity of the cold gas ($T<10^4 K$) at 2.2 disc scalelengths (8.4 kpc).
A single component Sersic fit to the profile of stellar mass surface density 
gives an index of $n=1.4$, close to an exponential profile, and an effective radius of 4.1 kpc. 
However, there is a clear up-turn in the profile inside the first kpc, so two exponential components provide a better fit. The resulted disc-only scale-lenght is 3.8 kpc, consistent with the observed size-velocity relations.
The relatively small central component has a mass of only 20\% of the total stellar mass, consistent with estimations from local SDSS galaxies. 
The fraction of light in the B-band is even smaller. Only 14\% of the luminosity in the B-band is coming from the first kpc.
Therefore, we conclude that the simulated galaxy is a low-mass, disc-dominated galaxy at $\simeq0$.

\section{Elongated Galaxies at Cosmic Noon ($z\simeq2$)}

\begin{figure}[b]
\begin{center}
 \includegraphics[width=3.4in]{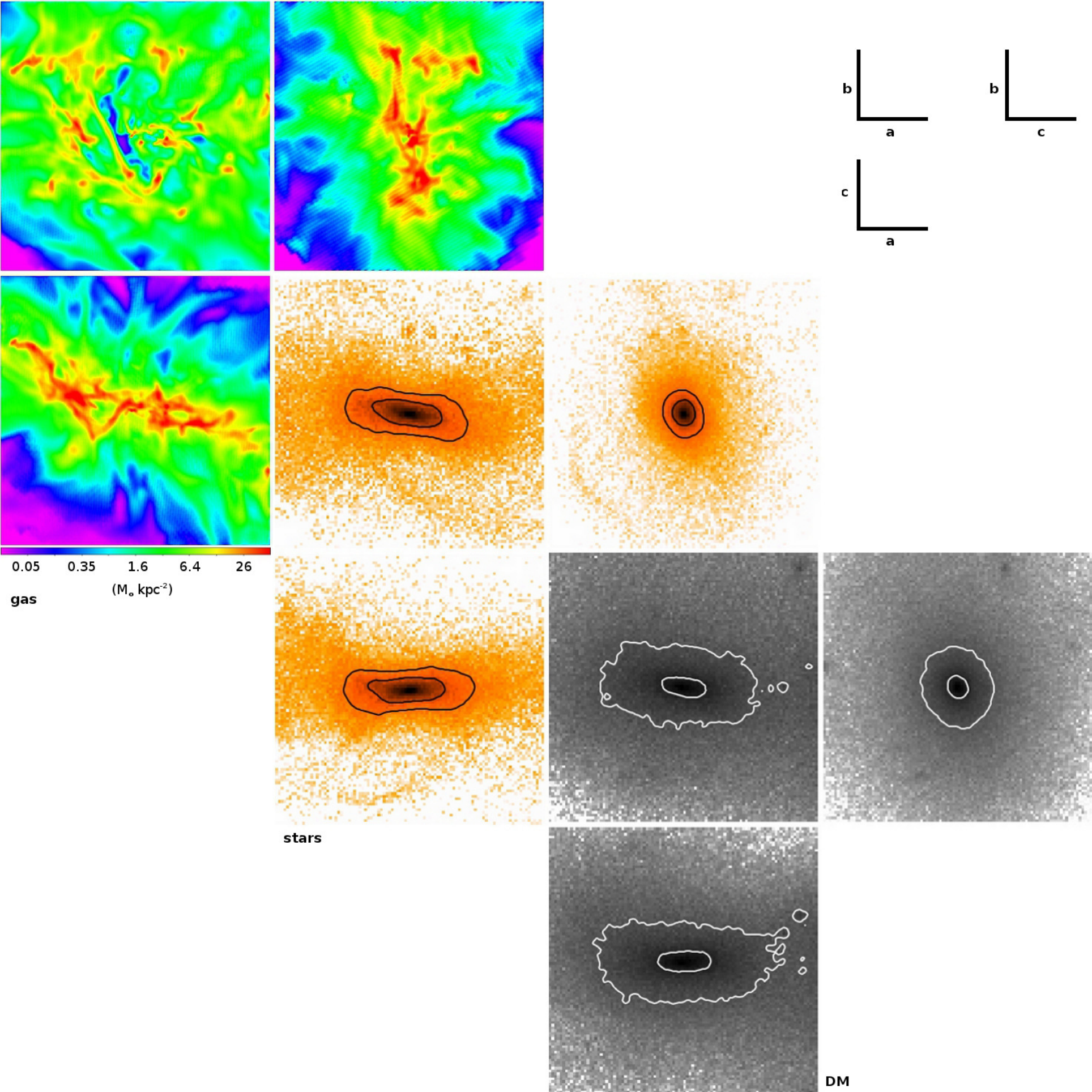} 
 \caption{Example of an elongated galaxy. Orthogonal views along the galaxy principal axes, showing gas, stars, and DM (from top to bottom) of Vela28 at $z=2.2$. The size of the images is 20 kpc. The contours of stellar surface density are defined as log($\Sigma/ (M_\odot {\rm kpc}^{-2}$)  =\ 1, 1.5 
in the b-a and c-a planes and  as log($\Sigma/ (M_\odot {\rm kpc}^{-2}$) =\ 1.5, 2 in the b-c plane. 
The stellar half-mass radius is $r_e=2.1 {\rm kpc}$. 
The contours of DM surface density are set as log($\Sigma/ (M_\odot {\rm kpc}^{-2}$) =\ 2, 2.5
along the minor and intermediate axes and as llog($\Sigma/ (M_\odot {\rm kpc}^{-2}$) =\ 2.2, 2.8 along the major axis.
The labels at the top-right corner mark the directions of the galaxy axes in each view.}
   \label{fig2}
\end{center}
\end{figure}

A typical galaxy with a stellar mass of around $10^9 \Msun$ at cosmic noon looks very different than the z=0 counterpart discussed above. 
Its effective radius is a factor 2 smaller (2 kpc) and its SFR is more than a factor 10 higher ($ {\rm SFR}=6 \sy$), reflecting the peak of the cosmic star formation history. 

The morphology of the galaxy (Vela28) is also strikingly different. It is not a disc or a spheroid but an elongated galaxy. 
Fig.\,\ref{fig2}  shows three orthogonal views of the three components (gas, stars, and DM) at $z=2.2$. The face-on view (b-a plane) was selected using the angular momentum of the cold ($T<10^4$ K) gas inside 5 kpc.
 The gas shows a very irregular and multiphase medium with clouds of relatively high column densities, log($\Sigma / \msun \ {\rm pc}^{-2} )=1.5-1.7$, and bubbles with very low densities. 
 The geometry is far from a uniform, thin disc, although the gas is flattened along a minor axis, $c$, coincident with the rotation axis of the gas. 
 Indeed, the gas rotates at ${\rm V}_{\rm rot}=80 \kms$, averaged over a 2-5 kpc shell. 
 The radial velocity dispersion measured at 200 pc scales is also high, $\sigma_{\rm r}=30 \kms$, typical of very perturbed, turbulent and thick discs. See \cite[Ceverino et al. (2012)]{Ceverino12} for more details.

The stellar distribution is very different from the thick gaseous disc. 
It is elongated along one direction, which defines the galaxy major axis $a$.
The 3D isodensity surface crossing the major axis around the half-mass radius is well fitted by an ellipsoid with axial lengths equal to $(a,b,c)=(2.2,0.76,0.67) \kpc$. 
The intermediate-to-major axis ratio is low, $b/a=0.35$, inconsistent with an axisymmetric disc. 
The shape is close to a prolate 
ellipsoid.

The dark matter also shows a prolate shape with similar properties at roughly the same radius, 
$(a,b,c)_{\rm DM}=(1.8,0.73,0.6) \kpc$ and $(b/a)_{\rm DM}=0.4$. 
The fact that the dark matter is also elongated in the same direction as the stellar component 
is relevant to the formation of such an elongated galaxy. 
Elongated galaxies form preferentially in highly elongated, low-mass haloes at high redshifts.
This is due to asymmetric accretion from narrow filaments and mergers.
At the same time, feedback prevents the overabundance of baryons at the centre of these small haloes.
As galaxies grow in mass, baryons start to dominate the central potential. This makes the inner halo rounder, probably due to the deflection of elongated orbits by a central mass concentration.
The process of compaction (\cite[Zolotov et al. 2015]{Zolotov}), in which a gas-rich galaxy shrinks into a compact star-forming spheroid, could drive the transition from a DM-dominated elongated galaxy to a baryonic self-gravitating spheroid.

\section{Dwarfs at Cosmic Dawn ($z=6-15$)}

\begin{figure}[b]
\begin{center}
 \includegraphics[width=3.4in]{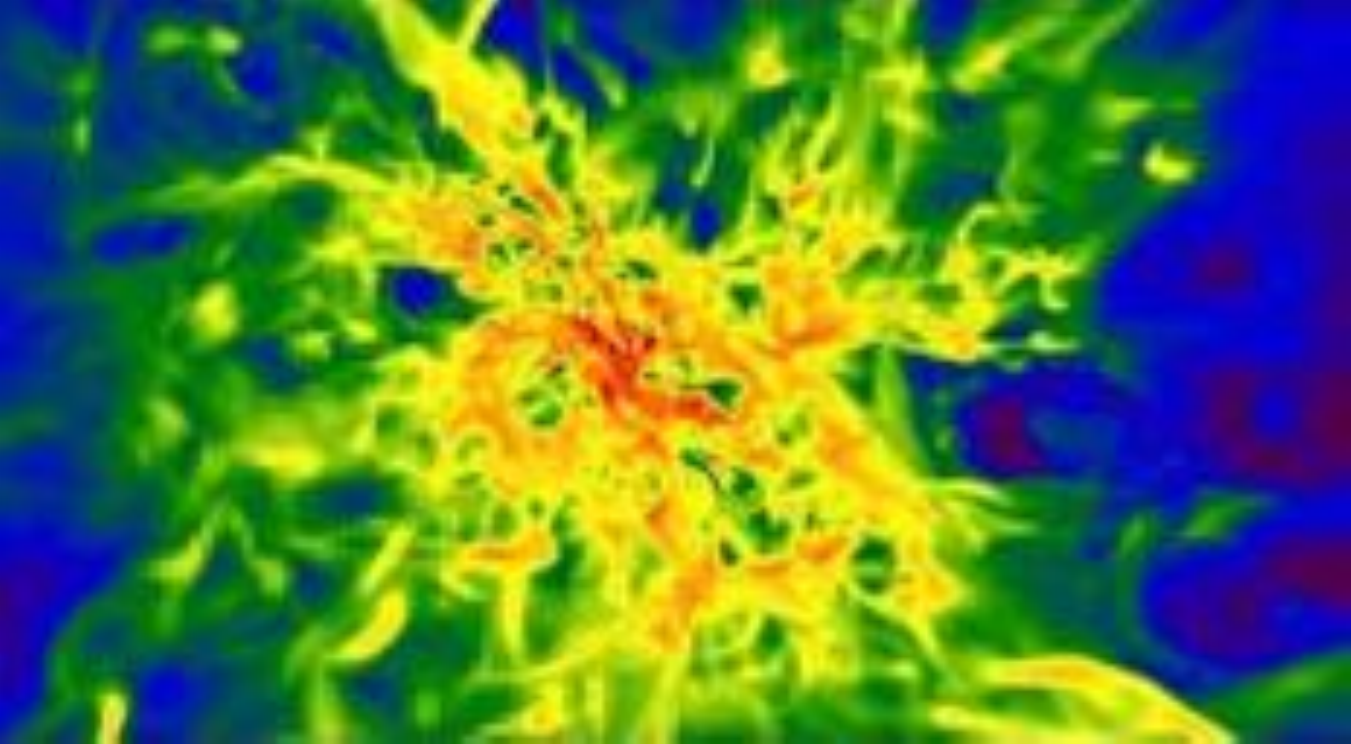} 
 \caption{Gas distribution of a dwarf galaxy at $z=6$. The horizontal size of the image is 4 kpc.}
    \label{fig3}
\end{center}
\end{figure}

Galaxies at the LMC mass scale at cosmic dawn look very compact (Fig.\,\ref{fig3}). Its stellar half-mass radius is only 0.5 kpc. This is a factor 10 smaller than at z=0. They are gas rich and they have typical $ {\rm SFR}\simeq20 \sy$, a factor 10 higher than their counterparts at $z=0$. This drives a clumpy and turbulence gas morphology, more similar to local starbursts. We have used the FirstLight database to study the Star Formation (SF) histories of $\sim$300 galaxies with a stellar mass between $\Ms=10^6$ and $3 \times 10^9 \Msun$ during cosmic dawn ($z=5-15$). 

The evolution of the SFR in each galaxy is complex and diverse, characterized by bursts of SF. 
Overall, first galaxies spend about 70\% of their time undergoing SF bursts at $z>5$.
This diversity sets the mean and scatter of the SFMS at $z\simeq5-13$.
High gas fractions and short gas depletion times are common during the SF bursts.
The typical bursts at $z\simeq6$ have a sSFR maximum of  $5-15 \GyrI$ with a duration of $\sim$100 Myr, one tenth of the age of the  Universe.
A quarter of the bursts populate a tail with very high sSFR maxima of $20-30 \GyrI$ and significantly shorter time-scales of $\sim 40-80$ Myr at all masses.
The mean period of time between consecutive bursts is $\sim$200 Myr with a small mass dependence at $z\simeq6$. 
The mean sSFR increases with redshift approximately as $sSFR \propto (1+z)^{5/2}$ at all masses, as predicted by $\Lambda$CDM models. This is consistent with existing observations at $z\leq8$.
The typical sSFR height of a SF burst also increases with redshift, but it is always a factor $2\pm0.5$ higher than the mean sSFR at that redshift.
This implies typical sSFR maxima of $sSFR_{\rm max}=20-30 \GyrI$ at $z=9-10$. The tail of the distribution reaches $sSFR_{\rm max}\simeq 60 \GyrI$ at these high redshifts.
This evolution is driven by shorter time-scales at higher redshifts, proportional to the age of the Universe.
These galaxies are the most efficient star formers in the history of the Universe.


\begin{thebibliography}{}

\bibitem[Ceverino et al. (2012)]{Ceverino12} 
{Ceverino D., Dekel A., Mandelker N., Bournaud F., Burkert A., Genzel R.,
Primack J.} 2012,  \textit{MNRAS}, 420, 3490

\bibitem[Ceverino, Primack, Dekel (2015)]{Ceverino15b}
{Ceverino D., Primack J., Dekel A.} 2015, \textit{MNRAS}, 453, 408

\bibitem[Ceverino et al. 2014]{Ceverino14}
{Ceverino D., Klypin A., Klimek E. S., Trujillo-Gomez S., Churchill C. W., Primack J., Dekel A.} 2014, \textit{MNRAS}, 442, 1545

\bibitem[Ceverino et al.(2017)]{Ceverino17} 
{Ceverino, D., Primack, J., Dekel, A., \& Kassin, S.A.} 2017,  \textit{MNRAS}, 467, 2664 

\bibitem[Ceverino, Glover, Klessen 2017]{PaperI}
{Ceverino D., Glover S. C. O., Klessen R. S.} 2017, \textit{MNRAS}, 470, 2791

\bibitem[Ceverino, Klessen, Glover (2018)]{PaperII}
{Ceverino D., Klessen R. S., Glover S. C. O.} 2018, \textit{MNRAS}, 480, 4842

\bibitem[Zolotov et al. 2015]{Zolotov}
{Zolotov A. et al.} 2015, \textit{MNRAS}, 450, 2327

\end{thebibliography}
\end{document}